\documentclass[fleqn,usenatbib]{mnras}

\usepackage{newtxtext,newtxmath}

\usepackage[T1]{fontenc}
\usepackage{ae,aecompl}

\usepackage{graphicx}	
\usepackage{amsmath}	
\usepackage{amssymb}	

\title[$^{13}$CO : N$_2$ and $^{13}$CO$_2$ : N$_2$ ice mixtures]{$^{13}$CO and $^{13}$CO$_2$ ice mixtures with N$_2$ in photon energy transfer studies}

\author[H. Carrascosa et al.]{
  H. Carrascosa,$^{1}$\thanks{E-mail: hcarrascosa@cab.inta-csic.es}
  L. -C. Hsiao,$^{2}$
  N. -E. Sie$^{2}$
  G. M. Mu\~noz Caro$^{1}$\thanks{E-mail: munozcg@cab.inta-csic.es}
  Y. -J. Chen$^{2}$\thanks{E-mail: asperchen@phy.ncu.edu.tw}
\\
$^{1}$Centro de Astrobiolog\'{\i}a (CSIC-INTA), Ctra. de Ajalvir, km 4, Torrej\'on de Ardoz, 28850 Madrid, Spain\\
$^{2}$Department of Physics, National Central University, Jhongli City, Taoyuan County 32054, Taiwan\\
}

\date{Accepted 2019 March 26. Received 2019 March 25; in original form 2019 January 30.}

\pubyear{2019}

\begin{document}
\label{firstpage}
\pagerange{\pageref{firstpage}--\pageref{lastpage}}
\maketitle

\begin{abstract}
In dense clouds of the interstellar medium, dust grains are covered by ice mantles, dominated by H$_2$O. CO and CO$_2$ are common ice components observed in infrared spectra, while infrared inactive N$_2$ is expected to be present in the ice. Molecules in the ice can be dissociated, react or desorb by exposure to secondary ultraviolet photons. Thus, different physical scenarios lead to different ice mantle compositions. This work aims to understand the behaviour of $^{13}$CO : N$_2$ and $^{13}$CO$_2$ : N$_2$ ice mixtures submitted to ultraviolet radiation in the laboratory. Photochemical processes and photodesorption were studied for various ratios of the ice components. Experiments were carried out under ultra-high vacuum conditions at 12K. Ices were irradiated with a continuous emission ultraviolet lamp simulating the secondary ultraviolet in dense interstellar clouds. During the irradiation periods, fourier-transform infrared spectroscopy was used for monitoring changes in the ice, and quadrupole mass spectrometry for gas-phase molecules. In irradiated $^{13}$CO$_2$ : N$_2$ ice mixtures, $^{13}$CO, $^{13}$CO$_2$, $^{13}$CO$_3$, O$_2$, and O$_3$ photoproducts were detected in the infrared spectra. N$_2$ molecules also take part in the photochemistry, and N-bearing molecules were also detected: NO, NO$_2$, N$_2$O, and N$_2$O$_4$. Photodesorption rates and their dependence on the presence of N$_2$ were also studied. As it was previously reported, $^{13}$CO and $^{13}$CO$_2$ molecules can transfer photon energy to N$_2$ molecules. As a result, $^{13}$CO and $^{13}$CO$_2$ photodesorption rates decrease as the fraction of N$_2$ increases, while N$_2$ photodesorption is enhanced with respect to the low UV-absorbing pure N$_2$ ice.
\end{abstract}




\section{Introduction}
In the Interstellar Medium (ISM), dense clouds are made of H$_2$ and other gases, while sub-micron dust grains contribute about 1\% of the cloud mass. As a consequence of the low temperatures, near 10 K in the cloud core, dust grains are covered by small molecules, leading to the formation of ice mantles. H$_2$O is the main constituent of the ice and smaller amounts of CO, CO$_2$, CH$_3$OH, CH$_4$, NH$_3$, etc, are commonly observed \citep{Mumma2011,Boogert2011}. Molecules with weaker intermolecular forces such as CO, and presumably N$_2$, are found in a larger extent on the surface of ice mantles, as they need lower temperatures to freeze out (T < 20 K) \citep{Pontoppidan2008}. In the cold astrophysical environments with temperatures below 20 K, non-thermal desorption processes induced by cosmic rays and secondary UV photons are the main processes contributing to the abundance of CO and N$_2$ gases.\\

Secondary UV radiation is generated by the interaction of cosmic rays with hydrogen molecules present in dense clouds \citep{Prasad1983,Cecchi1992,Shen2004}.  As a consequence, mainly ion and photon induced desorption are the processes expected to contribute to the observed gas phase abundances below 20 K. Ice molecules can desorb by two different processes induced by UV photons. Desorption Induced by Electronic Transition (DIET) occurs when a molecule absorbs a photon, and it experiments an electronic transition to the excited state. Then, it can follow two different routes. If the molecule is located on the surface of the ice, it can use this energy to desorb, what is known as direct DIET. If the molecule is in the subsurface monolayers (MLs), the electronic energy is distributed to the surrounding species. Molecules in the surface can receive this energy, which is used to break the intermolecular bonds and photodesorb \citep{Bertin2012,MD2015}. This process is known as indirect DIET. Direct DIET is negligible in comparison with indirect DIET, as most photons are absorbed in the subsurface MLs compared to the absorption in the first ML of the ice. Photochemical desorption or photochemidesorption is a different process. Some molecules do not desorb efficiently upon irradiation of the pure ice, but they photochemidesorb as photoproducts of ices with a different composition (\cite{MD2015} and references therein).\\

Desorption induced by secondary UV radiation is a complex process, it varies depending on the molecules present in the ice, the interactions between them and the ice temperature, among others. A few works were dedicated to study the photon-induced desorption in multicomponent ices \citep{Bertin2013, Fillion2014, MD2015, MD2016, Gus2016, Gus2018}. This work reports UV radiation of binary ices formed by $^{13}$CO and $^{13}$CO$_2$ mixed with N$_2$ were studied to explore the formation, photon-induced and thermal desorption of photoproducts. Irradiation experiments of ice mixtures with similar compositions were reported by \cite{SandfordAlla1990, Elsila1997} in high vacuum experiments, but, unfortunately, ultra-high vacuum (UHV) is required to study photon-induced desorption processes. N$_2$ has a very low vacuum ultraviolet (VUV) absorption cross section below 12 eV. Thus, photodesorption of pure N$_2$ ice is negligible in our experiments (see Sect. \ref{Experimental setup}). Keeping the same $^{13}$CO or $^{13}$CO$_2$ ice thickness, different deposition ratios with N$_2$ were used, to investigate the photodesorption yield of the molecules.\\

Apart from photodesorption, secondary UV radiation within interstellar dense clouds and cold circumstellar regions can induce destruction and formation of covalent bonds in the ice. Consequently, new compounds appear in the ice mantles. Different species have been detected in the ISM and protoplanetary disks, but their formation pathway is still unknown. The processes of formation of photoproducts can be elucidated by experimental simulations. In this study, photoproducts obtained from the different mixtures are also discussed.

\section{Experimental setup}
\label{Experimental setup}
The experiments were carried out using the UltraHigh Vacuum Interstellar Photoprocess System (UHV-IPS) at National Central University in Taiwan. The experimental protocol was described in \cite{Asper2014}. During the experiments, the system was cooled down to 12 K, and the base pressure was around 3$\cdot$10$^{-10}$ torr at room temperature. Fourier-Transform Infrared spectroscopy (FTIR) and Quadrupole mass spectrometry (QMS) data were taken before and after cooling down to cryogenic temperatures to check the presence of any potential contaminants. A T-type microwave discharge hydrogen flow lamp (MDHL) was used to simulate the background UV radiation in dense clouds. Light source stability of MDHL was also checked before starting the experiments. The deposition substrate was irradiated to quantify the blank signal in the QMS, to substract it from the ice signal desorption afterwards. Using a gas line system, the ice mixtures for each experiment were prepared. High purity gases were used: $^{13}$CO$_2$ (99\%), $^{13}$CO (99\%), N$_2$ (99.999\%). The gas flow was directed through stainless steel bellows, using a 1 mm diameter capillary to inject it into the chamber. The capillary was placed 20 mm away, aligned at normal angle of the CaF$_2$ substrate. Additionally, another MgF$_2$ window acted as interface between the MDHL and the chamber, leading to a cut-off at 114 nm (10.8 eV). The composition of the growing ice was monitored by a QMS, and the ice thickness with FTIR.\\

The ratio of the gas mixture components was measured before introducing them in the chamber. First, one of the components is introduced in a gas line system which has four bottles with the same volume, until it reached the desired pressure, and stored at the same pressure in an independent bottle. The gas line system is then evacuated. The same procedure is applied to the second gas component. Both gases are finally mixed in the gas line. However, the proportions obtained in the deposited ices were slightly different. The amount of each component was recalculated during deposition using the integrated area for the molecular $\frac{m}{z}$ fragments, giving the final composition (see Table \ref{Table1}) assuming that all molecular components share a similar sticky coefficient. This allowed to estimate the column density of IR-inactive N$_2$. All the experiments were repeated twice to ensure the results were reliable. During deposition of the ices, the column density of the components was controlled using FTIR, to obtain a value of 200 ML (1 ML is $10^{15}$ molecules cm$^{-2}$) of $^{13}$CO or $^{13}$CO$_2$.\\

The ices were processed by UV-radiation with the MDHL. A hydrogen flux of 0.4 mbar was fixed in the MDHL circuit. MDHL photon flux is determined in situ by a nickel mesh \citep{Asper2014}. After each irradiation period, the photon dose was  calculated, and an IR spectrum of the deposited ice was taken with a minimum resolution of 4 cm$^{-1}$. As the QMS spectrum was recorded during the whole irradiation sequence, both the solid and the gas phase were monitored.\\

\begin{table*} 
   \centering
      \caption[]{$^{13}$CO$\thinspace$:$\thinspace$N$_{2}$ and $^{13}$CO$_{2}\thinspace$:$\thinspace$N$_{2}$ ice mixture experiments performed for this study.}
    \label{Table1}
    \begin{tabular}{ccccccc}
Experiment	& Ratio $^{13}$CO:N$_{2}$*		&N $^{13}$CO (ML)**			&&Experiment		& Ratio $^{13}$CO$_{2}$:N$_{2}$*		&N $^{13}$CO$_{2}$ (ML)**\\
\noalign{\smallskip}
\cline{1-3}
\cline{5-7}
\noalign{\smallskip}
\cline{1-7}
\noalign{\smallskip}
1			&1:0					&200					&&7		    	&1:0					&196\\
\noalign{\smallskip}
2			&2:1					&201					&&8	    		&2:1					&201\\
\noalign{\smallskip}
3			&1:1					&201					&&9	    		&1:1					&201\\
\noalign{\smallskip}
4			&1:2					&198					&&10			&1:2					&199\\
\noalign{\smallskip}
5			&1:4					&199				    &&11			&1:4					&199\\
\noalign{\smallskip}
6			&0:1					&0\\ 			
\noalign{\smallskip}
\hline
\end{tabular}\\
\begin{flushleft}
\textit{
* Rounded values are given to facilitate the discussion. Error was below 5\% in all cases and the exact values were used for the graphs (e. g. the 1:1 ice mixture was 49\% of $^{13}$CO and 51\% of N$_2$).\\ 
** Average value between the two experiments was taken, the difference was below 6 ML in all cases.}\\
\end{flushleft}
\end{table*}

The ices were irradiated leading to total irradiation times of 5, 10, 15, 20, 25, 35, 45, 55, 70, and 90 min. From these data, ice column densities of $^{13}$CO$_2$ and $^{13}$CO were calculated from their infrared absorption band, using equation \ref{Eq.1} and integrating from 2320 cm$^{-1}$ to 2220 cm$^{-1}$ and from 2120 cm$^{-1}$ to 2075 cm$^{-1}$, respectively. In this formula, $N$ is the column density in molecules $\cdot$ cm$^{-2}$, $A$ is the band strength in cm $\cdot$ molecule$^{-1}$ (listed in Table \ref{Table5}), $\tau _{v}$ the optical depth of the band, and $dv$ the wavenumber differential in cm$^{-1}$.\\

\begin{equation}
\centering
\;\;\;N = \frac{1}{A} \int_{band}{\tau _{v} dv}
\label{Eq.1}
\end{equation}\\

From QMS data, the photodesorption rate (molecules per incident photon) was obtained. $\frac{m}{z} = 45$ was used to monitor $^{13}$CO$_2$, $\frac{m}{z} = 29$ for $^{13}$CO, $\frac{m}{z} = 28$ for N$_2$, $\frac{m}{z} = 32$ for O$_2$, and $\frac{m}{z} = 48$ for O$_3$. Blank irradiation of the substrate with no ice and baseline corrections were made to obtain reliable data. The area of the QMS signal during each irradiation period was integrated for each of the $\frac{m}{z}$ values. The accumulated QMS area was represented as a function of photon dose and fitted linearly using the last 4 values, where the slope is constant. In this way, the ion current per number of incident photons was obtained. To convert it into the number of molecules of each species, equation \ref{Eq.2} was applied, following \citet{MD2015}.\\

\begin{eqnarray}
 \begin{split}
      \;\;\;N(mol)\; = \;\;& A\left(\frac{m}{z}\right) \;\;x\;\; k_{CO} \;\;x\;\; \frac{\sigma^{+}(CO)}{\sigma^{+}(mol)} \;\;x\;\;
      \frac{IF(CO^{+})}{IF(z)} \;\;x \\
      &\\
      &x\;\; \frac{FF(28)}{FF(m)} \;\;x\;\; \frac{S(28)}{S(\frac{m}{z})} \;\;x\;\; \frac{1x10^{15} mol}{1 ML}
  \label{Eq.2}
 \end{split}
\end{eqnarray}\\
 
In this formula, $A$($\frac{m}{z}$) is the integrated area below the QMS signal, $k_{CO}$ is a proportionality constant which depends on the QMS and the set-up, calculated for CO as the reference (for more details, see \citet{MD2015}), $\sigma^{+}$ is the ionization cross section, both for the sample molecule and the reference, IF is the ionization factor, FF the fragmentation factor, S the sensitivity of the QMS and the last term converts MLs into number of molecules. For calculation purposes, the values adopted are the ones shown in Table \ref{Table2}. $\frac{IF(CO^{+})}{IF(z)}$ was considered unity in all cases, as most of the molecules are ionized once.\\

\begin{table} 
   \centering
   \caption[]{Values adopted for QMS calculation}
   \label{Table2}
   \begin{tabular}{cccccc}
Factor			& N$_2$			&$^{12}$CO			&$^{13}$CO		&O$_2$		&$^{13}$CO$_2$\\
\noalign{\smallskip}
\hline
\hline
\noalign{\smallskip}
\noalign{\smallskip}
k\thinspace\thinspace($\frac{ML}{A \cdot scan}$)&-			&$5.26 \cdot 10^{9}$	&-			&-		&-\\
\noalign{\smallskip}
$\sigma^{+}$ (mol $\cdot$ \AA$^{2}$)*&2.508	&2.516			&2.516			&2.441		&3.521\\
\noalign{\smallskip}
FF				&0.892		&0.93			&0.93			&0.661		&0.69\\
\noalign{\smallskip}
S ($\frac{m}{z} \cdot \thinspace 10^2)$		&3.120          &3.120  	&2.957  	&2.517 	&1.418\\
\noalign{\smallskip}
\noalign{\smallskip}
\hline
\noalign{\smallskip}
\end{tabular}
\begin{flushleft}
\textit{
* Data were taken from the National Institute of Standards (NIST) with an estimated value between 5-20\%}
\end{flushleft}
\end{table}

\section{Results and discussion}
This section is divided into three different subsections. Sect. \ref{13CO.mixtures} introduces $^{13}$CO : N$_2$ mixtures, as a prerequisite to interpret $^{13}$CO$_2$ : N$_2$ mixtures. The experiments are presented in Sect. \ref{13CO2.mixtures}. A comparison of the results using both ice mixtures is provided in Sect. \ref{Comparison}.\\

\subsection{$^{13}$CO : N$_2$ results}
\label{13CO.mixtures}

\begin{figure*} 
  \centering
  \includegraphics[width=0.65\textwidth]{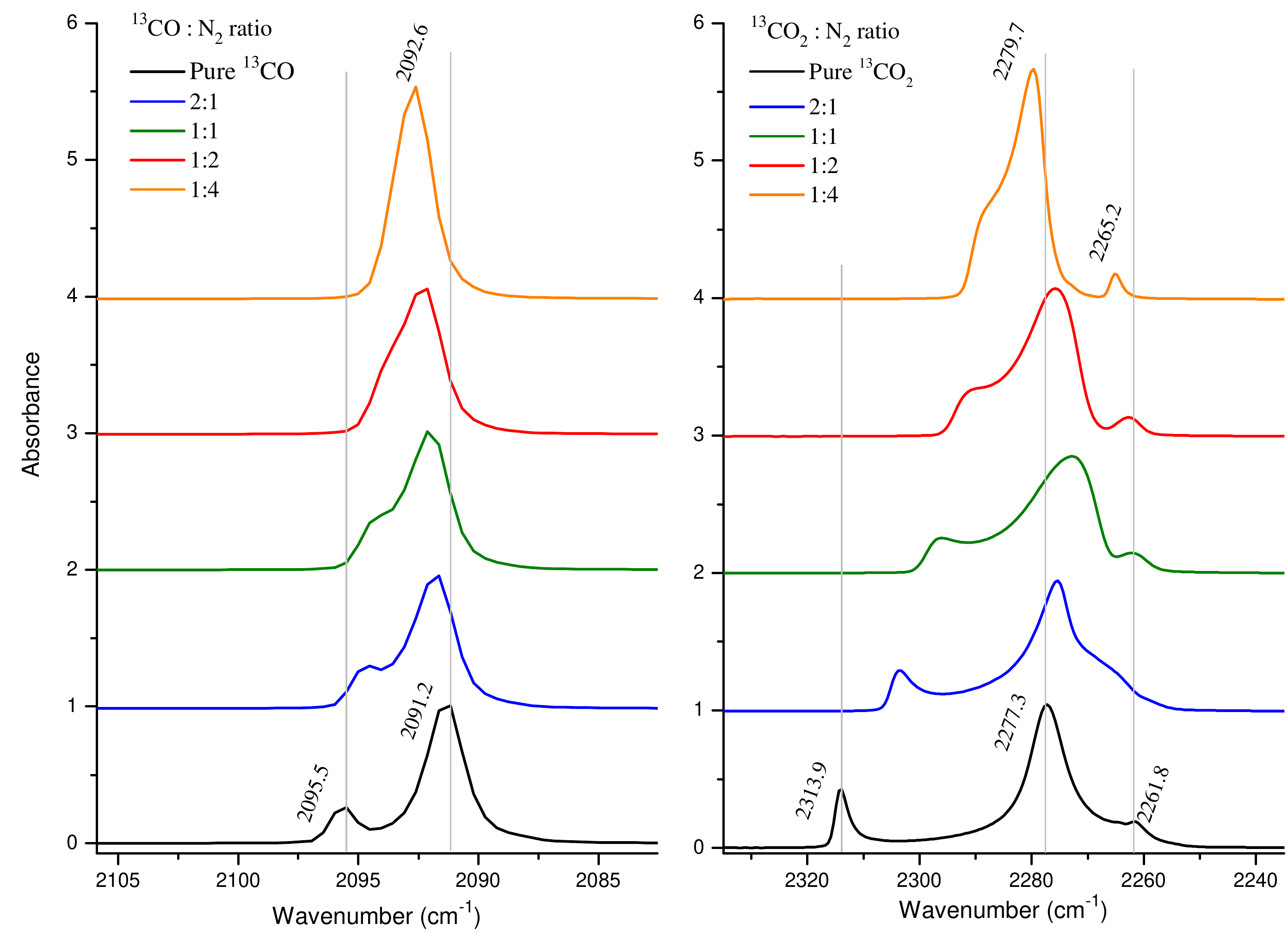}
  \caption{Comparison of IR spectra (resolution of 1 cm$^{-1}$) of the deposited ices, $^{13}$CO : N$_2$ ice mixtures on the right and $^{13}$CO$_2$ : N$_2$ ice mixtures on the left.}
  \label{Fig.2}
\end{figure*}

Figure \ref{Fig.2} shows the spectra before irradiation for the different ice mixtures. Two peaks appear in the pure $^{13}$CO ice spectrum measured at 45$^{\circ}$ incidence angle, which converge as more N$_2$ is added to the mixture. Thus, $^{13}$C$\equiv$O stretching is splitted into two bands: one band belongs to the transverse optical (TO) mode and another one to the longitudinal optical (LO) mode, centered at 2091.2 cm$^{-1}$ and 2095.5 cm$^{-1}$, respectively \citep{Palumbo2006}. As the amount of N$_2$ is increased, the shape of the bands changes. TO mode is redshifted while LO is blueshifted. When the fraction of N$_2$ exceeds 0.66, both bands overlap and become indistinguishable. In other words, the LO and TO modes characteristic of cubic $^{13}$CO ice vanish when there is enough N$_2$ in the ice mixture, as it might be expected. Additionally, the IR band of $^{13}$CO becomes narrower as the proportion of N$_2$ is increased. N$_2$ acts similar to noble gases used for dilution of ice molecules, therefore reducing intermolecular van der Waals forces between $^{13}$CO molecules.\\

\begin{table} 
    \caption[]{Band strength values adopted for column density calculations.}
    \label{Table5}
    \begin{center}
    \begin{tabular}{ccc}
Species		&Frequency  &Band strength\\
\noalign{\smallskip}
&(cm$^{-1}$)    &(cm molecule$^{-1}$)\\
\noalign{\smallskip}
\hline
\hline
\noalign{\smallskip}
\noalign{\smallskip}
$^{13}$CO	&2093	&1.3 $\cdot$ 10$^{-17}$ $^{a}$\\
\noalign{\smallskip}
$^{13}$CO$_2$	&2280   &7.8 $\cdot$ 10$^{-17}$ $^{a}$\\
\noalign{\smallskip}
$^{13}$CO$_3$	&1989  &1.5 $\cdot$ 10$^{-17}$ $^{b}$\\
\noalign{\smallskip}
O$_3$	&1041	&1.4 $\cdot$ 10$^{-17}$ $^{c}$\\
\noalign{\smallskip}
NO	&1875   &4.5 $\cdot$ 10$^{-18}$ $^{d}$\\
\noalign{\smallskip}
N$_2$O	&2235   &6.1 $\cdot$ 10$^{-17}$ $^{e}$\\
\noalign{\smallskip}
NO$_2$  &1614   &6.2 $\cdot$ 10$^{-17}$ $^{e}$\\
\noalign{\smallskip}
\noalign{\smallskip}
\hline
\noalign{\smallskip}
\end{tabular}
\end{center}
\textit{
$^{a}$ From \cite{Gerakines1995}\\
$^{b}$ Value taken from $^{12}$CO$_3$ in \cite{MD2015}\\
$^{c}$ From \cite{Smith1985}\\
$^{d}$ From \cite{Sicilia2012}\\
$^{e}$ From \cite{Fulvio2009}\\}
\end{table}

\begin{figure*} 
  \centering 
  \includegraphics[width=0.9\textwidth]{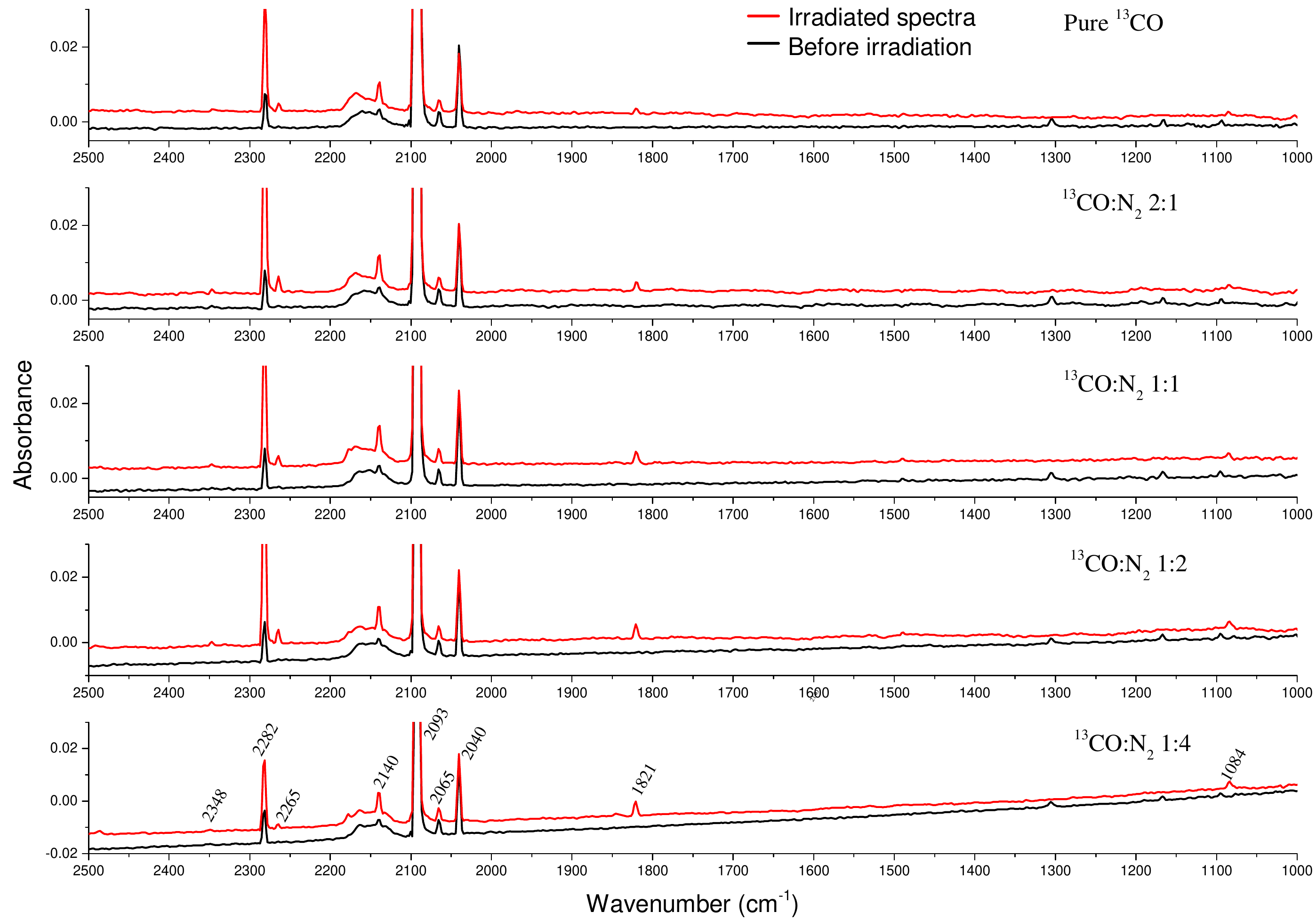}
  \caption{IR spectra before and after 90 min irradiation for the different $^{13}$CO : N$_2$ mixtures.}
  \label{Fig.7}
\end{figure*}

Fig. \ref{Fig.7} and Table \ref{Table4} show the IR features for the different $^{13}$CO : N$_2$ ice mixtures. Direct $^{13}$CO dissociation requires an energy up to 11.1 eV (112 nm) that is not attained in our experiments using the MDHL with a MgF$_2$ window interface. However, $^{13}$CO$_2$ molecules can be produced triggered by an excited state of $^{13}$CO, as follows:

\begin{eqnarray}
  \label{scheme1}
  \begin{split} 
  \centering
   \;\;\;&^{13}CO \thinspace + \thinspace h\nu \quad \;\;\;\;\;\:\! \xrightarrow{} \quad ^{13}CO\thinspace^{*}\\
   \;\;\;&^{13}CO^{*} \thinspace+ \thinspace ^{13}CO \quad \xrightarrow{} \quad ^{13}CO_2 \thinspace +\thinspace  ^{13}C\thinspace \cdot\\
  \end{split}
\end{eqnarray}

On the other hand, $^{13}$CO$_2$ dissociation energy is lower (5.44 eV, 228 nm) \citep{huebner1992}, and becomes efficient in these experiments. As a consequence, the decrease of the pure $^{13}$CO ice band due to the formation of $^{13}$CO$_2$ is < 3\%, and CO photodesorption consumes an important fraction of the absorbed photon energy during continued irradiation \citep{Guille2010}.

\begin{eqnarray}
  \label{scheme2}
  \begin{split} 
  \centering
  \;\;\;^{13}CO_2 \thinspace + \thinspace h\nu \quad &\xrightarrow{} \quad ^{13}CO \thinspace + \thinspace O \thinspace \cdot\\
   \end{split}
\end{eqnarray}

\begin{figure*} 
  \centering 
  \includegraphics[width=0.8\textwidth]{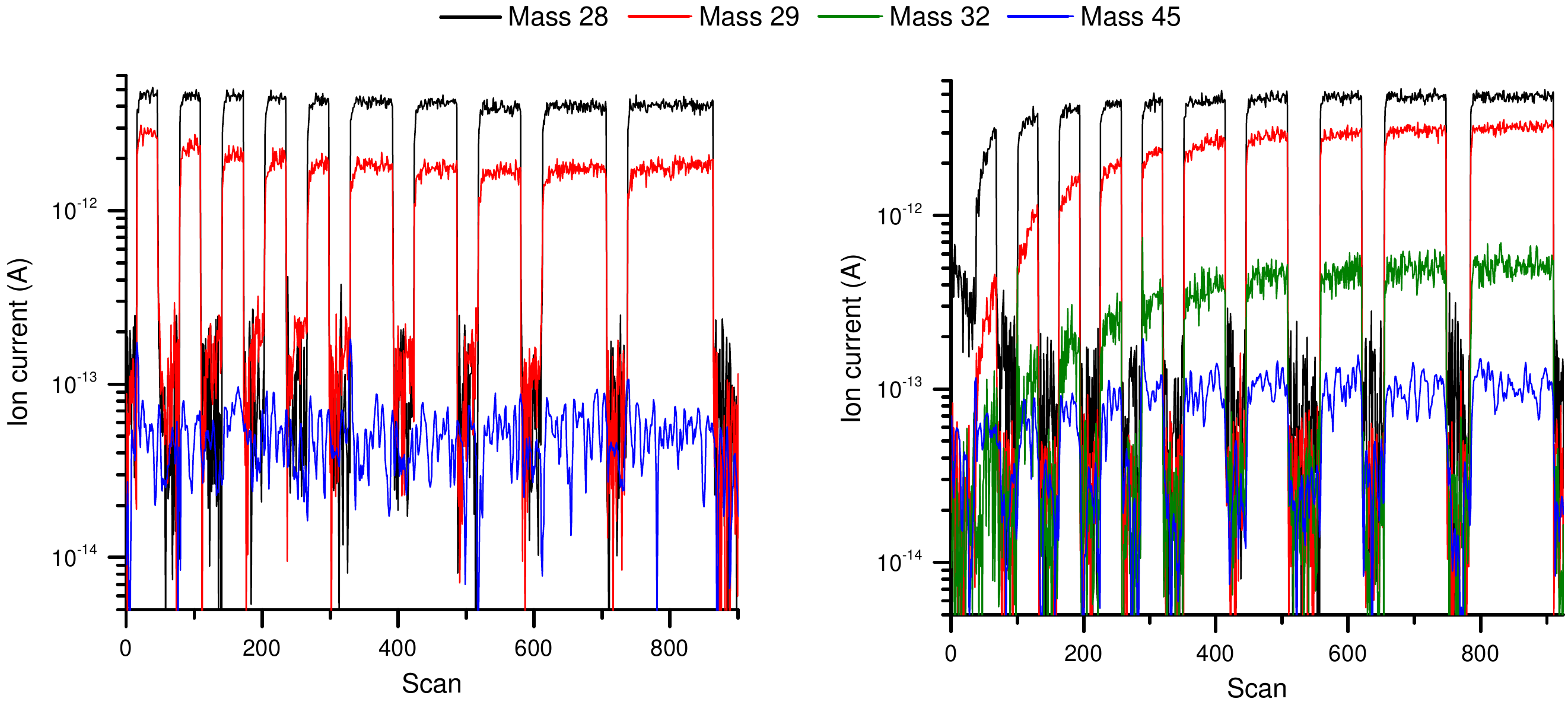}
  \caption{Ion current obtained for three different $\frac{m}{z}$ fragments during the irradiation of a $^{13}$CO : N$_2$ = 1:2 mixture in Experiment 4 (left) and four fragments from a $^{13}$CO$_2$ : N$_2$ = 1:1 mixture in Experiment 9 (right), see Table 1. A 5-point median smooth was applied to $\frac{m}{z}=45$ in both graphs for clarity.}
  \label{Fig.3}
\end{figure*} 

\begin{table*} 
   \centering
    \caption[]{IR bands (cm$^{-1}$) detected after 90 min irradiation for different $^{13}$CO:$\thinspace$N$_{2}$ mixtures, experiments 1-5.}
    \label{Table4}
    \resizebox{19cm}{!} {
    \begin{tabular}{ccccccc}
Pure $^{13}$CO		&$^{13}$CO:N$_{2}$ (2:1)		&$^{13}$CO:N$_{2}$ (1:1)			&$^{13}$CO:N$_{2}$ (1:2)		&$^{13}$CO:N$_{2}$ (1:4)		&Assignment & References\\
\noalign{\smallskip}
\hline
\hline
\noalign{\smallskip}
\noalign{\smallskip}
2345	&2347	&2347	&2347	&2348	&$^{12}$CO$_2$  &\cite{Yamada1964}\\
\noalign{\smallskip}
2281	&2282	&2282	&2282	&2282	&$^{13}$CO$_2$  &\cite{SandfordAlla1990}\\
\noalign{\smallskip}
2264	&2264	&2264	&2265	&2265	&$^{13}$C$^{18}$O$^{16}$O  &This work\\
\noalign{\smallskip}
2140	&2140	&2140	&2140	&2140	&$^{12}$CO  &\cite{Jiang1975}\\
\noalign{\smallskip}
2092	&2092	&2092	&2093	&2093	&$^{13}$CO  &\cite{Gerakines1995}\\
\noalign{\smallskip}
2065    &2065	&2065	&2065	&2065	&$^{13}$C$^{17}$O  &\cite{Loeffler2005}\\
\noalign{\smallskip}
2040    &2040	&2040	&2040	&2040	&$^{13}$C$^{18}$O  &\cite{Oberg2009}\\
\noalign{\smallskip}
1820	&1820	&1820	&1820	&1821	&?  &\\
\noalign{\smallskip}
-   	&-  	&-  	&1085	&1084	&?  &\\
\noalign{\smallskip}
\noalign{\smallskip}
\hline
\noalign{\smallskip}
\end{tabular}
}
\begin{center}
\textit{}
\end{center}
\end{table*}

\begin{figure} 
  \centering 
  \includegraphics[width=0.4\textwidth]{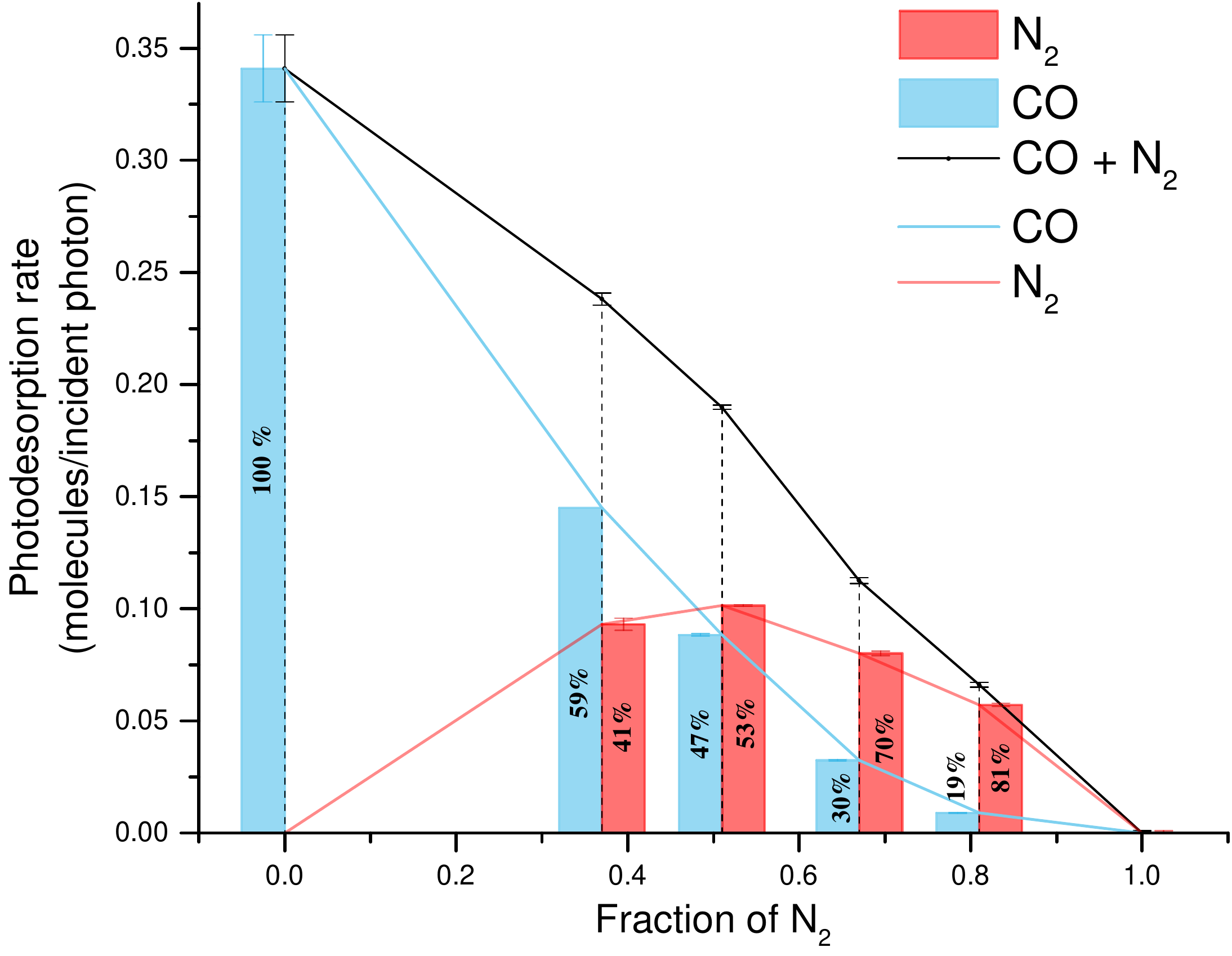}
  \caption{Photodesorption rate per incident photon for CO, N$_2$ and CO$+$N$_2$ in $^{13}$CO : N$_2$ mixtures.}
  \label{Fig.5}
\end{figure}

From QMS measurements during ice deposition, it was possible to calculate the $^{13}$CO : N$_2$ ratio, obtaining the values shown in Table \ref{Table1}, which are slightly different from the ones measured in the gas line prior to deposition (see Sect. \ref{Experimental setup}). QMS data (Fig. \ref{Fig.3}) allowed estimation of photodesorption ratios for each experiment. The results are shown in Figure \ref{Fig.5}. $^{13}$CO is not able to desorb directly from its own fragmentation in the excited state, as it is a non-dissociative state \citep{Okabe1978,Cottin2003}. Thus, $^{13}$CO desorbs mainly through an indirect DIET mechanism.\\

As \cite{Asper2017} reported, N$_2$ and $^{13}$CO energy levels only overlap in the low wavelength emission range of the MDHL. N$_2$ ice VUV absorption ranges from 120 nm to 145 nm, and $^{13}$CO ice absorbs photons from 130 nm to 160 nm. From 120 nm to 130 nm, their absorption is relatively low. \cite{Bertin2013} studied this effect, concluding that direct energy transfer is barely possible between both molecules. Energy transfer follows an indirect DIET mechanism. Photoabsorption induces a transition from the electronic ground state to the first excited one. Energy from the relaxation of $^{13}$CO to the ground state can reach a N$_2$ molecule on the ice surface, where conversion to translational energy can lead to desorption.\\

The N$_2$ photodesorption rate increases as a function of the fraction of N$_2$ up to a maximum, for a fraction of N$_2$ of 0.5 (red line in Fig. \ref{Fig.5}). Simultaneously, $^{13}$CO photodesorption rate decreases (blue line). Interestingly, the CO$+$N$_2$ photodesorption rate follows a linear trend; its slope is related to the number of photons absorbed by $^{13}$CO, since N$_2$ absorption is negligible. Additionally, Fig. \ref{Fig.5} shows that both CO and N$_2$ have the same possibilities to desorb (e. g. a fraction of N$_2$ of 0.8 is traduced into a 81\% of N$_2$ photodesorption and 19\% of CO photodesorption, and so on for the other ratios). Knowing that, only, the top 5$\pm$1 MLs contribute to the photodesorption of pure CO ice \citep{Guille2010, Fayolle2011,Asper2014}, two different effects lead to the observed values. First, as the proportion of N$_2$ is increased, there will be fewer $^{13}$CO molecules and more N$_2$ molecules on the ice surface ready to photodesorb. Second, the lower number of $^{13}$CO molecules on the top 5$\pm$1 MLs reduces the number of absorbed photons that contribute to photodesorption. Thus, the energy available for photodesorption decreases and the overall value of photodesorption is reduced (black line).\\

\subsection{$^{13}$CO$_2$ : N$_2$ ice mixtures}
\label{13CO2.mixtures}

In Figure \ref{Fig.2} various $^{13}$CO$_2$ : N$_2$ ice spectra after deposition are shown. The profile of the $^{13}$C=O stretching band of $^{13}$CO$_2$ is altered by the presence of N$_2$. In a way analog to $^{13}$CO : N$_2$ ice mixtures, two bands, belonging to the TO and LO modes at 2277 cm$^{-1}$ and 2314 cm$^{-1}$, respectively, are seen in the pure $^{13}$CO$_2$ ice, and they converge into one band for large N$_2$ ratios, although a shoulder in the band shows that it is not a single isolated peak.\\

Table \ref{Table3} shows the different IR features after 90 min irradiation for different $^{13}$CO$_2$ : N$_2$ mixtures. N$_2$ plays a role in the photoprocessing of the ices. It is expected that the presence of N$_2$ between $^{13}$CO$_2$ molecules would result in lower molecular interactions. However, $^{13}$CO$_2$ : N$_2$ intermolecular interactions induce the formation of other photoproducts (see Table \ref{Table5}). $^{13}$CO is the main photoproduct, as it is formed easily from the rupture of $^{13}$CO$_2$ molecules (Scheme \ref{scheme2}).\\


O$_2$ and O$_3$ are also produced from UV irradiation of $^{13}$CO$_2$ : N$_2$ mixtures (see scheme \ref{scheme3} and Table \ref{Table3}). However, O$_2$ has no dipole moment, neither intrinsic or induced by IR photons. Thus, its presence was confirmed by QMS ($\frac{m}{z} = 32$) during the irradiation of the ices and temperature programmed desorption (TPD) experiments, while O$_3$ was identified from IR spectra.

\begin{eqnarray}
  \label{scheme3}
  \begin{split} 
  \centering
   \;\;\;&O \thinspace \cdot \thinspace + \thinspace O \thinspace \cdot \thinspace \quad \xrightarrow{} \quad O_2\\
   \;\;\;&O_2 \thinspace + \thinspace h\nu \quad \;\xrightarrow{} \quad O \cdot \thinspace + \thinspace O \thinspace \cdot\\
   \;\;\;&O_2 \thinspace + \thinspace O \thinspace \cdot \thinspace \quad \xrightarrow{} \quad O_3\\
   \;\;\;&O_3 \thinspace + \thinspace h\nu \quad \;\xrightarrow{} \quad O_2 \thinspace + \thinspace O \thinspace \cdot\\
 \end{split}
\end{eqnarray}

$^{13}$CO$_3$  is formed as shown in Scheme \ref{scheme4}. $^{13}$CO$_3$ abundance decreases as more N$_2$ is introduced in the ice mixture, as a consequence of the lower number of intermolecular $^{13}$CO$_2$ - $^{13}$CO$_2$ interactions. As QMS was recorded from $\frac{m}{z} = 1$ to $\frac{m}{z} = 50$, $^{13}$CO$_3$ was only detected by an infrared absorption band at 1989 cm$^{-1}$, and confirmed by the position and evolution of this band. The IR band position looks coherent with the bathochromic shift observed for $^{13}$CO$_2$ and $^{13}$CO from the corresponding $^{12}$CO$_2$ and $^{12}$CO molecules (see Table \ref{Table4}). Additionally, $^{13}$CO$_3$ is formed at short irradiation times and its abundance decreases for longer irradiation periods \citep{MD2015}.

\begin{figure*} 
  \centering 
  \includegraphics[width=0.9\textwidth]{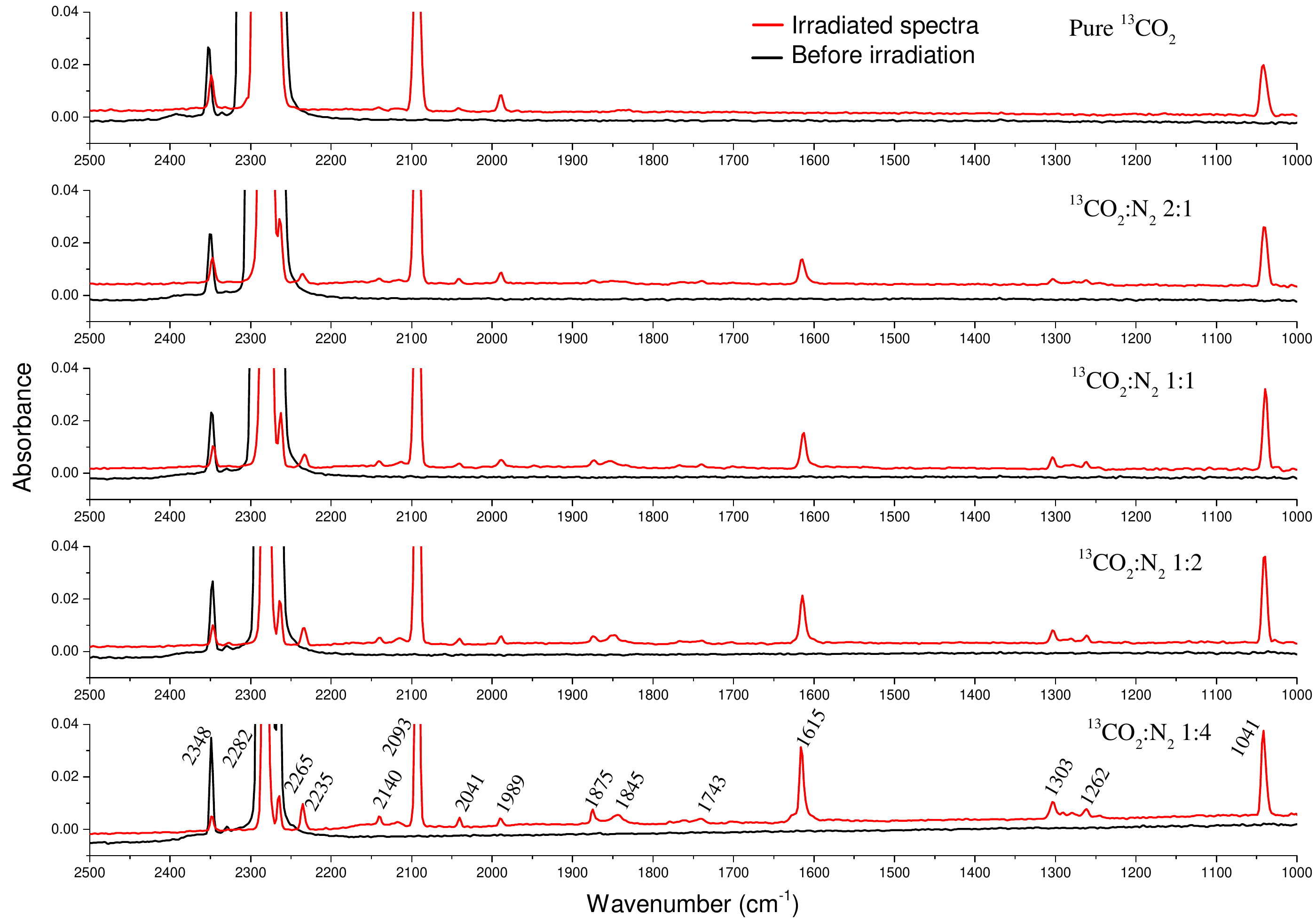}
  \caption{IR spectra before and after 90 min irradiation for $^{13}$CO$_2$ : N$_2$ ice mixtures.}
  \label{Fig.8}
\end{figure*}

\begin{eqnarray}
  \label{scheme4}
  \begin{split} 
  \centering
   \;\;\;&^{13}CO_2 \thinspace + \thinspace O \thinspace \cdot \thinspace \quad \xrightarrow{} \quad ^{13}CO_3\\
   \;\;\;&^{13}CO_3 \thinspace + \thinspace h\nu \quad \;\xrightarrow{} \quad ^{13}CO_2 \thinspace + \thinspace O \thinspace \cdot\\
  \end{split}
\end{eqnarray}

\begin{figure} 
  \centering 
  \includegraphics[width=0.4\textwidth]{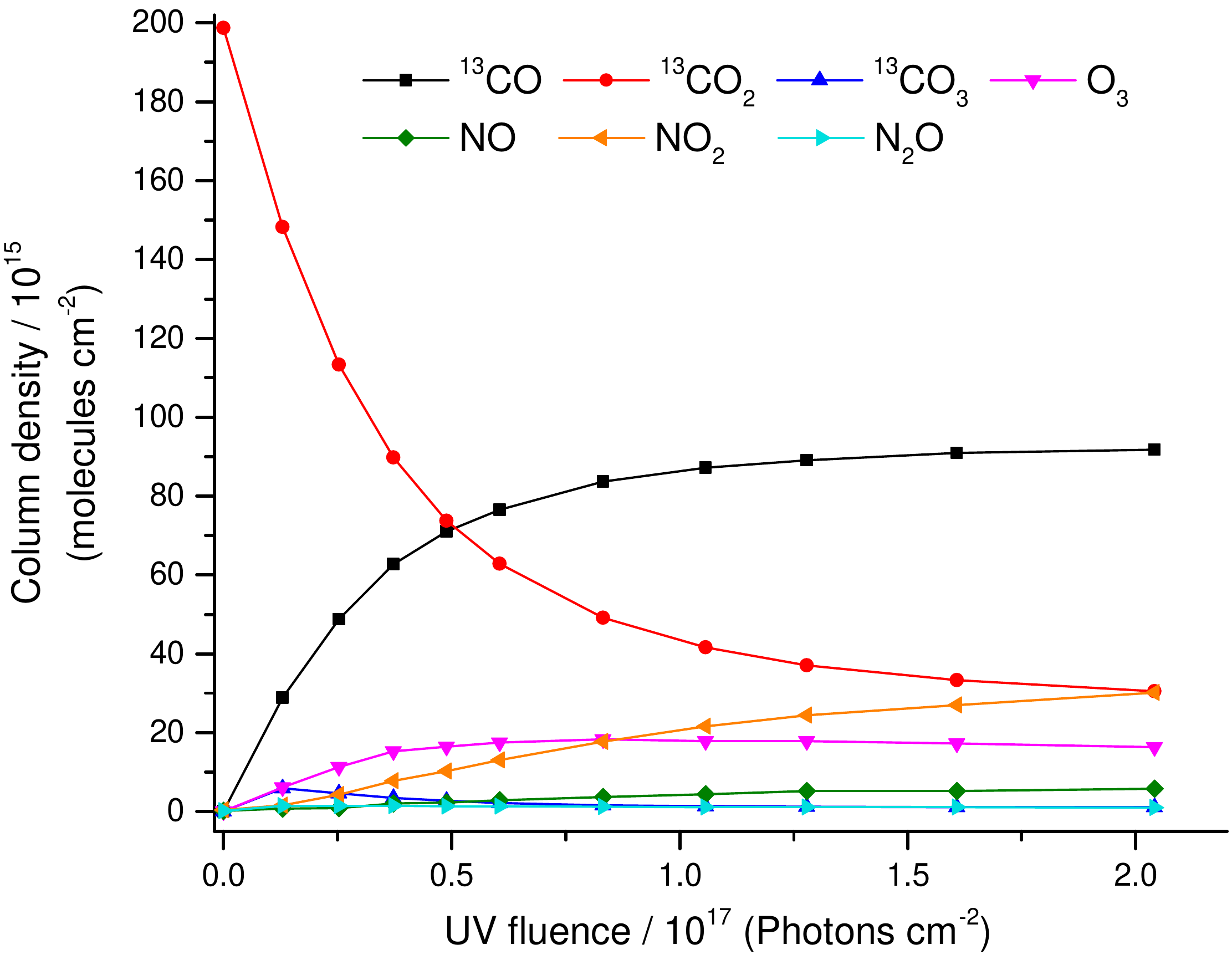}
 \caption{Evolution of the column densities of $^{13}$CO$_2$ and its photoproducts (Exp. 11) as a function of the photon dose using the band strength values shown in Table \ref{Table5}.}
  \label{Fig.9}
\end{figure}

These photoproducts had also been detected in pure CO$_2$ ices in previous works \citep{Oberg2009,Bahr2012,MD2015}. Additionally, up to four different N-bearing molecules formed from $^{13}$CO$_2$ - N$_2$ interactions, have been detected in this work: NO, N$_2$O, NO$_2$ and N$_2$O$_4$. Their abundances as a function of photon dose (molecules per induced photon) are represented in Fig. \ref{Fig.9}. N$_2$O is supposed to be formed easily, as its formation does not require N$_2$ dissociation (see Scheme \ref{scheme5}). However, it is not the most abundant species found.

\begin{eqnarray}
  \label{scheme5}
  \begin{split} 
  \centering
   \;\;\;&N_2 \thinspace + \thinspace O \thinspace \cdot \thinspace \quad \;\;\xrightarrow{} \quad N_2O\\
   \;\;\;&N_2O \thinspace + \thinspace h\nu \thinspace \quad \xrightarrow{} \quad N_2 \thinspace + \thinspace O \thinspace \cdot\\
   \end{split}
\end{eqnarray}

Other N$_2$ oxides can only be produced after dissociation of N$_2$ molecules, which implies a photon energy of 9.74 eV (127 nm), covered by the MDHL emission spectrum.

\begin{eqnarray}
  \label{scheme6}
  \begin{split} 
  \centering
   \;\;\;&N_2 \thinspace + h\nu \quad \;\;\xrightarrow{} \quad N \thinspace \cdot \thinspace + \thinspace N \thinspace \cdot\\
   \;\;\;&N \thinspace \cdot \thinspace + \thinspace N \thinspace \cdot \thinspace \quad \xrightarrow{} \quad N_2\\
   \end{split}
\end{eqnarray}

From N radicals, other nitrogen oxides can be formed.

\begin{eqnarray}
  \label{scheme7}
  \begin{split}
   \;\;\;&N \thinspace \cdot \thinspace + \thinspace O \thinspace \cdot \thinspace \quad \;\;\:\;\;\;\!\xrightarrow{} \quad NO\\
   \;\;\;&NO \thinspace + \thinspace h\nu \thinspace \quad \;\;\;\;\:\xrightarrow{} \quad N \thinspace \cdot \thinspace + \thinspace O \thinspace \cdot\\
   \;\;\;&N \thinspace \cdot \thinspace + \thinspace O_2 \quad \;\;\;\;\;\;\xrightarrow{} \quad NO_2\\
   \;\;\;&NO_2 + \thinspace h\nu \thinspace \quad \;\;\;\;\xrightarrow{} \quad N \thinspace \cdot \thinspace + \thinspace O_2\\
   \;\;\;&NO_2 \thinspace + \thinspace NO_2 \quad \:\xrightarrow{} \quad N_2O_4\\
   \;\;\;&N_2O_4 + \thinspace h\nu \thinspace \quad \;\;\:\xrightarrow{} \quad NO_2 \thinspace + \thinspace NO_2 \\
  \end{split}
\end{eqnarray}

During UV irradiation, nitrogen oxides can be dissociated. NO is the most stable one with a dissociation energy of 6.46 eV (191 nm), while NO$_2$ requires 3.11 eV (399 nm) and N$_2$O 3.63 eV (341 nm) \citep{huebner1992}.\\

\begin{figure} 
  \centering 
  \includegraphics[width=0.4\textwidth]{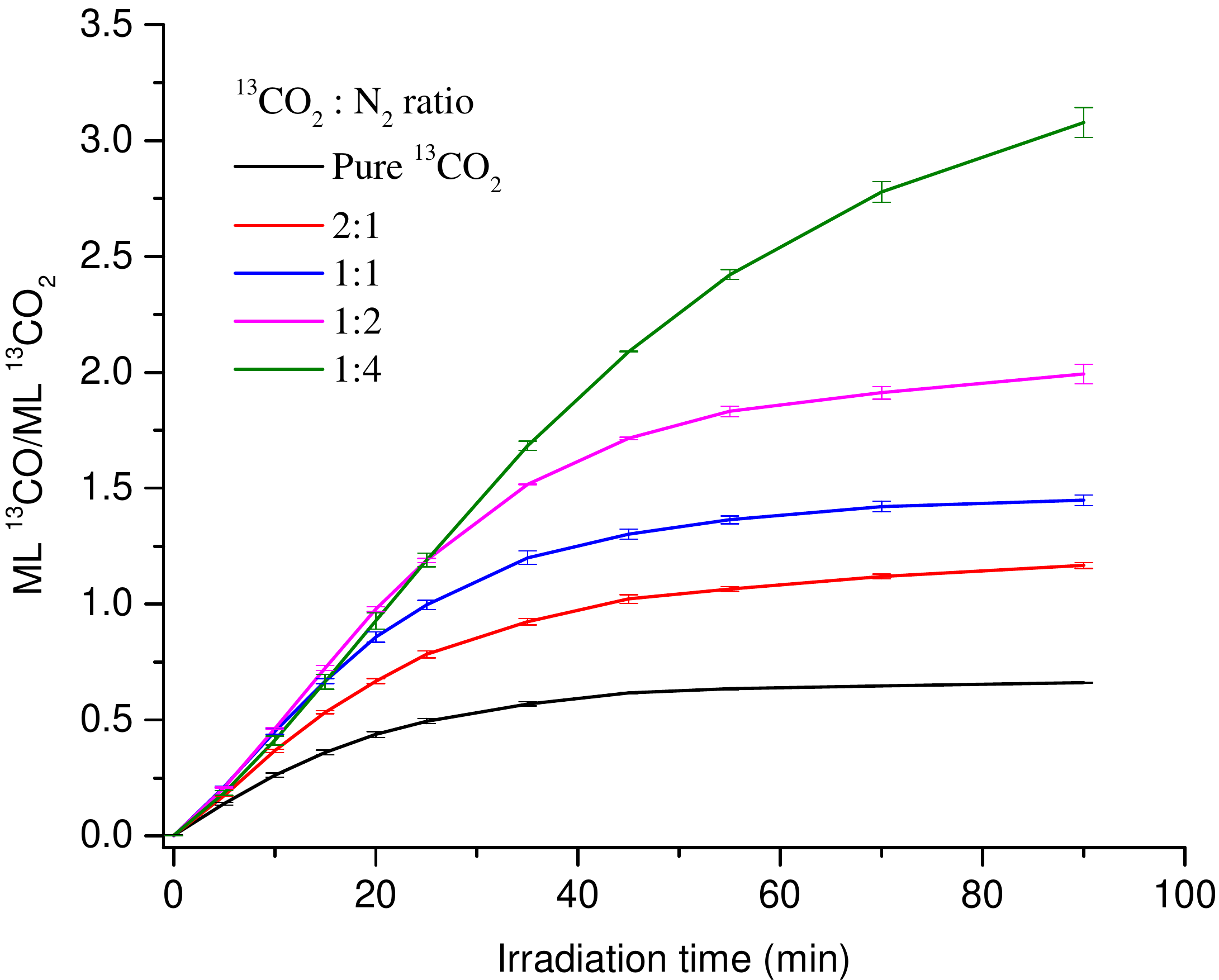}
  \caption{Ratio of the $\frac{^{13}CO}{^{13}CO_2}$ column densities during irradiation obtained from integration of the IR bands using Formula \ref{Eq.1}.}
  \label{Fig.4}
\end{figure}

\begin{table*} 
   \centering
    \caption[]{IR bands (cm$^{-1}$) detected after 90 min irradiations for different $^{13}$CO$_{2}:\thinspace$N$_{2}$ mixtures, experiments 7-11.}
    \label{Table3}
    \resizebox{19cm}{!} {
    \begin{tabular}{ccccccc}
Pure $^{13}$CO$_2$		&$^{13}$CO$_{2}$:N$_{2}$ (2:1)		&$^{13}$CO$_{2}$:N$_{2}$ (1:1)			&$^{13}$CO$_{2}$:N$_{2}$ (1:2)		&$^{13}$CO$_{2}$:N$_{2}$ (1:4)		&Assignment & References\\
\noalign{\smallskip}
\hline
\hline
\noalign{\smallskip}
\noalign{\smallskip}
3630	&3629	&3627	&3629	&3631	&$^{13}$CO$_2$  &This work\\
\noalign{\smallskip}
3513	&3514	&3514	&3516	&3618	&$^{13}$CO$_2$  &This work\\
\noalign{\smallskip}
2349    &2347	&2346	&2347	&2348	&$^{12}$CO$_2$  &\cite{Yamada1964}\\
\noalign{\smallskip}
2294    &2287	&-		&-		&-		&$^{13}$CO$_2$  &\cite{Gerakines1995}\\
\noalign{\smallskip}
2276	&2279	&2280	&2281	&2282	&$^{13}$CO$_2$  &\cite{SandfordAlla1990}\\
\noalign{\smallskip}
2264	&2264	&2263	&2264	&2265	&$^{13}$C$^{18}$O$^{16}$O   &This work\\
\noalign{\smallskip}
-    	&2236	&2233	&2234	&2235	&N$_2$O         &\cite{SandfordAlla1990,Fulvio2009}\\
\noalign{\smallskip}
2141	&2141	&2141	&2140	&2140	&$^{12}$CO  &\cite{Jiang1975}\\
\noalign{\smallskip}
2093	&2093	&2093	&2093	&2093	&$^{13}$CO  &\cite{Hudgins1993,Gerakines1996}\\
\noalign{\smallskip}
2042	&2041	&2041	&2041	&2041	&$^{13}$C$^{18}$O  &\cite{Oberg2009}\\
\noalign{\smallskip}
1989	&1989	&1989	&1989	&1989	&$^{13}$CO$_3$ &This work\\
\noalign{\smallskip}
-	    &1875	&1873	&1873	&1875	&NO        &\cite{Pugh1976, Sicilia2012}\\
\noalign{\smallskip}
-    	&-   	&1854	&1850	&1845	&NO/N$_2$O$_4$    &\cite{Fulvio2009}\\
\noalign{\smallskip}
-    	&-   	&1739	&1740	&1743	&N$_2$O$_4$    &\cite{Fulvio2009}\\
\noalign{\smallskip}
-    	&1614	&1613	&1614	&1615	&NO$_2$    &\cite{Fulvio2009}\\
\noalign{\smallskip}
-    	&1302	&1303	&1303	&1303	&NO$_2$/N$_2$O  &\cite{Fulvio2009}\\
\noalign{\smallskip}
-	    &1262	&1262	&1261	&1262	&N$_2$O$_4$    &\cite{Fulvio2009}\\
\noalign{\smallskip}
1041	&1040	&1039	&1041	&1041	&O$_3$  &\cite{Sicilia2012}\\
\noalign{\smallskip}
\noalign{\smallskip}
\hline
\noalign{\smallskip}
\end{tabular}
}
\begin{center}
\textit{}
\end{center}
\end{table*}

TPD data from QMS were used to confirm the presence of NO, N$_2$O and NO$_2$. The thermal desorption temperature of these photoproducts found in our experiments is coherent with previous works where these species were warmed up from pure ices \citep{Fulvio2009}.\\

\begin{figure} 
  \centering 
  \includegraphics[width=0.4\textwidth]{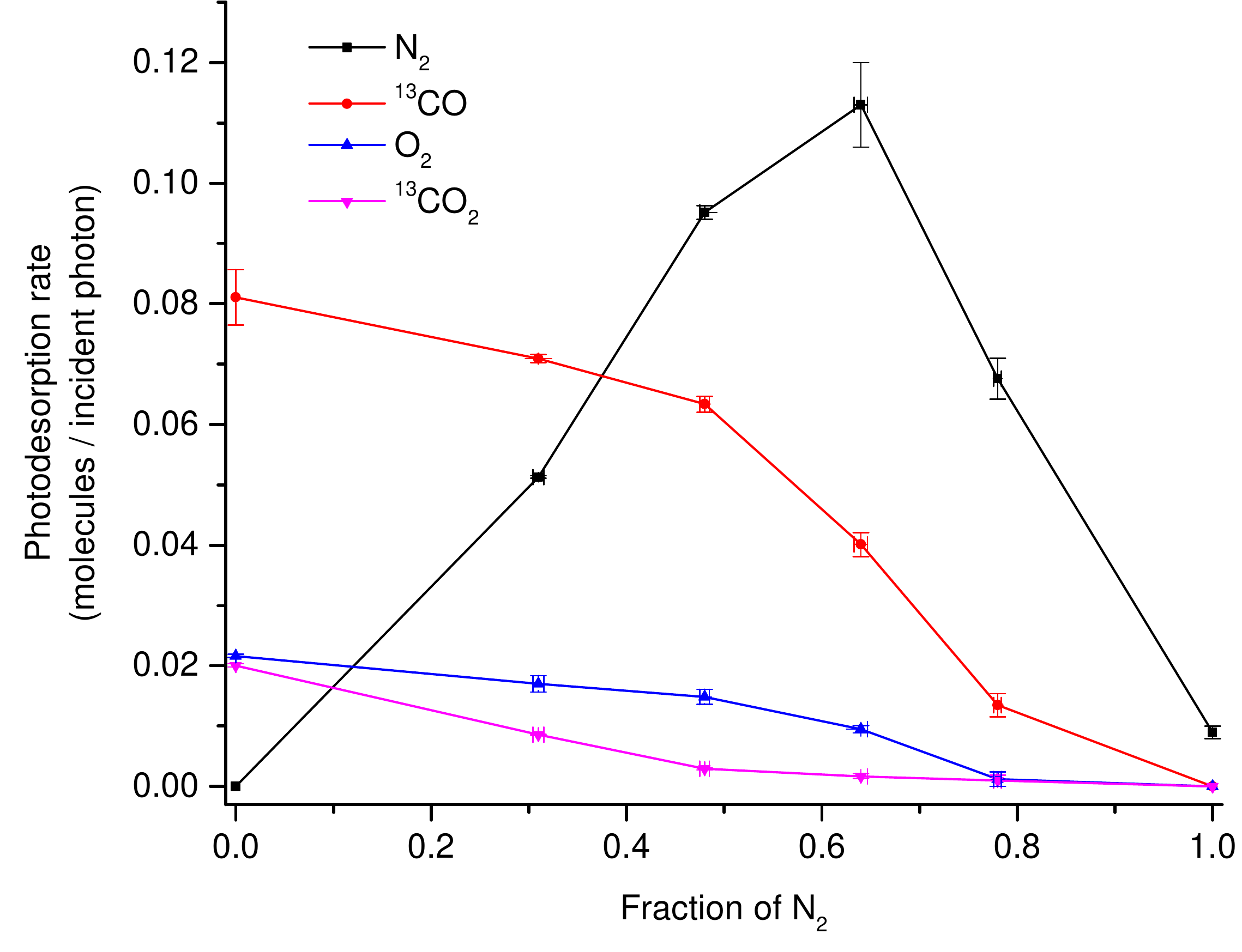} 
  \caption{Photodesorption obtained for $^{13}$CO$_2$ experiments as a function of the fraction of N$_2$.}
  \label{Fig.1}
\end{figure}

The QMS ion current measured during irradiation is related to the photodesorption rate, see Sect. \ref{Experimental setup}. The highest amount of photodesorbing N$_2$ was reached at the end of the irradiation. QMS data shows a correlation between $^{13}$CO and N$_2$ photodesorption. Their photodesorption rates increase during irradiation, what implies an energy transfer from $^{13}$CO to the UV low-absorbing N$_2$ molecules in the ice \citep{Bertin2013}. Although no photodissociation in pure N$_2$ ice, using the MDHL, was observed \citep{Gus2014}, irradiation of the $^{13}$CO$_2$ : N$_2$ mixtures lead to formation of NO, N$_2$, NO$_2$ and N$_2$O$_4$, detected by FTIR, (see Fig. \ref{Fig.8} and Table \ref{Table3}). Figure \ref{Fig.1} shows the photodesorption rate of the molecules of interest regarding $^{13}$CO$_2$ : N$_2$ mixtures. The $^{13}$CO photodesorption rate is higher than that of $^{13}$CO$_2$, even in the pure $^{13}$CO$_2$ ice. Photon-induced desorption of a specific nitrogen oxide molecule could not be confirmed in our experiments. The ion current of fragment NO$^{+}$ $\left(\frac{m}{z}=30\right)$ showed a significant increase upon irradiation, indicative of a photon-induced desorption. However, this fragment is common to various species (NO, NO$_2$, N$_2$O and N$_2$O$_4$).\\

Reaction to form $^{13}$CO from $^{13}$CO$_2$ is not unidirectional (see Schemes \ref{scheme1} and \ref{scheme2}), leading to a constant $\frac{^{13}CO}{^{13}CO_2}$ ice abundance ratio during irradiation. However, adding N$_2$ to $^{13}$CO$_2$ ices changes the equilibrium state. When a photon induces dissociation of a $^{13}$CO$_2$ molecule, the remaining photofragments will be placed further away from each other if more N$_2$ is present. Consequently, it becomes more difficult for two $^{13}$CO molecules to reform $^{13}$CO$_2$. The more N$_2$ in the sample, the larger amount of $^{13}$CO is present in the ice at the end of the irradiation period, as it is shown in Figure \ref{Fig.4}. As the amount of $^{13}$CO in the original $^{13}$CO$_2$ ice increases, the VUV absorption of the ice is shifted to longer wavelengths, which means that the emission spectrum of the MDHL lamp, in particular the Ly-$\alpha$ to molecular lines ratio, influences the outcome of these experiments. Therefore, in terms of UV absorption, upon irradiation, these ice mixtures can no longer be considered as binary $^{13}$CO$_2$ : N$_2$ mixtures, but ternary $^{13}$CO$_2$ : $^{13}$CO : N$_2$ mixtures, with variable ratios along the irradiation periods.\\

Pure N$_2$ ice photodesorption rate was below the QMS detection limit, as expected, due to its low VUV absorbance \citep{Gus2014b}. However, when N$_2$ is mixed with $^{13}$CO$_2$, N$_2$ photodesorption is observed. Thus, photon energy is transferred from the ice molecules with a significant absorbance in the VUV, mainly $^{13}$CO$_2$ and $^{13}$CO. Due to the cut off of the MgF$_2$ window at 114 nm, the main absorption bands of $^{13}$CO$_2$ play only a minor role, since only the tail of this band overlaps with the photon emission spectrum of the MDHL. On the other hand, the "A-X" absorption bands of $^{13}$CO match the emission range of the MDHL and contribute to the desorption of $^{13}$CO, N$_2$ and O$_2$. This is observed in Fig. \ref{Fig.3}, where photodesorption of these species goes by the hand. Additionally, Fig. \ref{Fig.1} shows that $^{13}$CO and O$_2$ photodesorption rates decrease at a similar rate for increasing fraction of N$_2$.\\

During the irradiation period, $^{13}$CO$_3$, $^{13}$CO$_2$, $^{13}$CO, O$_2$, O$_3$ and N$_2$ will be present in the ice matrix. Their abundances in the gas phase were monitored by QMS to quantify their photodesorption rates within the different experiments. As it is shown in Fig. \ref{Fig.1}, $^{13}$CO$_2$ photodesorption rate decreases readily as soon as N$_2$ is added to the ice mixture, indicative of an efficient energy transfer between both molecules. For larger N$_2$ ratios, indirect DIET from $^{13}$CO to N$_2$ is the dominant process, leading to a fast decrease of the $^{13}$CO photodesorption rate. \\

\subsection{Comparison between $^{13}$CO : N$_2$ and $^{13}$CO$_2$ : N$_2$ mixtures}
\label{Comparison}

\begin{figure} 
  \centering 
  \includegraphics[width=0.4\textwidth]{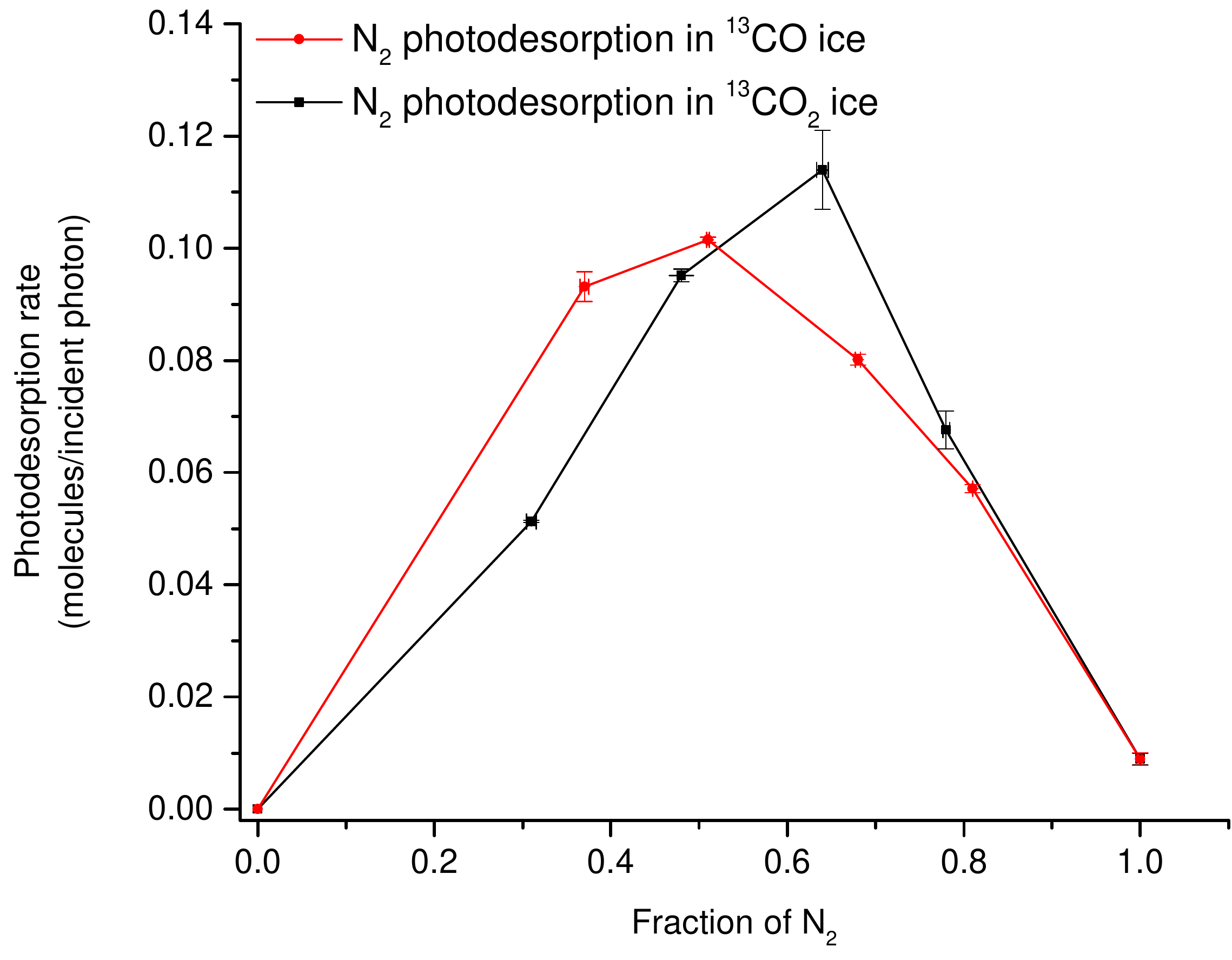}
  \caption{N$_2$ photodesorption rates in $^{13}$CO : N$_2$ and $^{13}$CO$_2$ : N$_2$ ice mixtures.}
  \label{Fig.6}
\end{figure} 

N$_2$ photodesorption rate as a function of the fraction of N$_2$ in $^{13}$CO and $^{13}$CO$_2$ ice is shown in Figure \ref{Fig.6}. The maximum photodesorption rate is different for both sets of mixtures. Regarding $^{13}$CO : N$_2$ mixtures, the maximum in the N$_2$ photodesorption is observed for a 0.5 fraction of N$_2$. However, for $^{13}$CO$_2$ : N$_2$ mixtures, the maximum is obtained for a fraction of N$_2$ around 0.64.\\

Regarding $^{13}$CO : N$_2$ mixtures, up to a fraction of N$_2$ of 0.5, N$_2$ photodesorption rate increases due to photon energy transfer from $^{13}$CO to N$_2$ in the ice. For fractions of N$_2$ above 0.5, N$_2$ photodesorption rate decreases. The number of absorbed photons in the top MLs is reduced, as a consequence of the lower number of $^{13}$CO molecules. Therefore, they cannot distribute their energy to all the N$_2$ molecules in the ice.\\

Concerning $^{13}$CO$_2$ : N$_2$ mixtures, the efficient formation of $^{13}$CO (Fig. \ref{Fig.4}), led to a higher VUV absorption cross section than pure $^{13}$CO$_2$ ice \citep{Gus2014}. VUV absorbance shifts to longer wavelengths as the ice is irradiated, contributing to N$_2$ photodesorption, as it occurs in the $^{13}$CO : N$_2$ mixtures. MDHL emitted VUV photons are absorbed in a broader wavelength range, while $^{13}$CO$_2$ in the ice absorbs from 114 nm to 140 nm, $^{13}$CO molecules absorb mainly from 140 nm to 162 nm. The addition of N$_2$ in the $^{13}$CO$_2$ ice obviously increases N$_2$ photodesorption, since more N$_2$ molecules are available at the ice surface, but the maximum does not occur at a fraction of N$_2$ of 0.5. Instead, it corresponds to a fraction of 0.64. The presence of N$_2$ molecules promotes $^{13}$CO formation from $^{13}$CO$_2$ (Fig. \ref{Fig.4}), and, as explained above, the combined absorption of both $^{13}$CO and $^{13}$CO$_2$ on the ice surface covers a broader photon energy range, thus enhancing photodesorption.\\

$^{13}$CO$_2$ ice photodissociates readily to form $^{13}$CO molecules. On the other hand, $^{13}$CO irradiation using T-type MDHL only produces a small amount of $^{13}$CO$_2$ (e. g. \cite{Guille2010,Asper2014}). This low conversion of $^{13}$CO$_2$ diminishes when $^{13}$CO is mixed with N$_2$. $^{13}$CO molecules absorb photons that induce mainly photodesorption rather than photochemistry. This is due to weak intermolecular forces and the low efficiency of $^{13}$CO dissociation. In these experiments, on the opposite, $^{13}$CO$_2$ absorbed photons are employed mainly in photochemistry, promoting the formation of photoproducts, such as $^{13}$CO$_3$, O$_2$, O$_3$, or nitrogen oxides.\\

\section{Astrophysical implications and conclusions}
The composition of ice mantles onto dust grains depends on the accretion scenario. Classical model \citep{Shu1977,Shu1987} predicts a constant accretion rate of molecules onto the dust grains. According to this model, luminosity and temperature in protostars vary slowly with time. Icy mantles are formed by species which do not receive enough energy to desorb thermally. Molecules such as H$_2$O and CH$_3$OH are detected on the warmest parts of the cloud, while CO and N$_2$ will be present in the ice mantles located in the dense clouds interior where the temperature is around 10 K. More recently, \cite{Dunham2010} proposed that protostars could experiment episodes of high and low luminosity. The variation of luminosity is translated into a different composition for ice mantles. During low luminosity periods, different molecules will be incorporated onto dust grains (e. g. CO). Then, secondary UV radiation induces chemical changes (e. g. formation of CO$_2$ from CO). High luminosity results in a higher temperature, when volatile species will thermally desorb. For instance, CO would thermally desorb, leaving CO$_2$ behind \citep{Kim2012, Dunham2010}.\\

CO : N$_2$ mixtures are present within the first scenario and low luminosity periods from the second scenario, while pure CO$_2$ : N$_2$ mixtures will not be present in astrophysical environments. However, CO : CO$_2$ : N$_2$ mixtures will be present within the episode accretion scenario. The ratio between the three components will vary depending on the luminosity period. As luminosity and temperature start to raise up, CO and N$_2$ will thermally desorb at a different rate depending on the environment.\\

This work is focused on scenarios at low temperatures, where CO and N$_2$ will form the top MLs of the ice. Thus, studying the photochemistry of ice mixtures including CO, CO$_2$ and N$_2$ turns out to be relevant to understand the behaviour of the dense clouds interior. Photodesorption rates of $^{13}$CO : N$_2$ and $^{13}$CO$_2$ : N$_2$ ice mixtures have been studied, as well as their dependence on the amount of N$_2$. $^{13}$CO absorbs photons and release energy to the surrounding molecules via indirect DIET. Therefore, the presence of N$_2$ diminishes the $^{13}$CO photodesorption rate. Fig \ref{Fig.5} shows that the ice composition determines the ratio between CO and N$_2$ photodesorption rates, while the absolute value of photodesorption depends on the absorbed photons. Concerning $^{13}$CO$_2$ : N$_2$ ice mixtures, direct energy transfer also plays a role, as $^{13}$CO$_2$ VUV absorption bands overlap with those of N$_2$. In both cases, N$_2$ photodesorption rate increases up to a maximum, as a consequence of the energy received from surrounding $^{13}$CO and, at a lesser extent, $^{13}$CO$_2$ molecules. For $^{13}$CO and $^{13}$CO$_2$ ice mixtures, fractions of N$_2$ above 0.5 and 0.64, respectively, imply a decrease in the photodesorption rate. This reduction is a result of the lower VUV absorbance in the top MLs.\\

In relation to photochemistry, pure $^{13}$CO and $^{13}$CO$_2$ ices have been extensively studied (\cite{MD2015,Ciaravella2016}, and references therein). However, a more realistic scenario including N$_2$ was not sufficiently explored regarding product formation and desorption. Nitrogen oxides have been detected in dense molecular clouds \citep{Ziurys1994,Halfen2001}, which may appear after dissociation of N$_2$ molecules and subsequent chemical reactions (see Schemes \ref{scheme6} and \ref{scheme7}). A reaction pathway for the formation of nitrogen oxides is proposed. $^{13}$CO$_2$ dissociation produces C and O radicals, which may react with N atoms or N$_2$ molecules to form NO, N$_2$O, NO$_2$ and N$_2$O$_4$ in the ice. Molecules carrying CN groups are also expected from reactions involving carbon radicals, although they have not been detected in this work.\\

CO, CO$_2$, N$_2$ or O$_2$ gas phase abundances at low temperatures in dense clouds can be estimated from the photodesorption rates calculated in this work. Photodesorption of photoproducts, such as $^{13}$CO$_3$, O$_3$, NO, N$_2$O, NO$_2$ and N$_2$O$_4$ was negligible at this low temperature. However, as the temperature is increased, nitrogen oxides co-desorb thermally with N$_2$, contributing to the gas phase abundance in dense clouds. Gas phase abundances can be explained, to a certain extent, based on the column density found in our IR spectra in agreement with the work of \cite{Elsila1997}. To look further, different formation pathways must be considered. Regarding nitrogen oxides, NO$_2$ was the most abundant species, followed by NO and N$_2$O. Observational measurements reported by \cite{Halfen2001} calculated the abundance of these molecules in Sagittarius B2 complex. They found NO column density (1.32 $\cdot 10^{16}$ cm$^{-2}$) to be larger than that of NO$_2$ (1.50 $\cdot 10^{15}$ cm$^{-2}$) and NO$_2$ (< 3.3 $\cdot 10^{15}$ cm$^{-2}$), suggesting that NO must be formed by other synthetic routes (e. g. involving H$_2$O molecules as a source of oxygen atoms). \cite{Halfen2001} suggested chemical reactions which may formed N$_2$O from NO or NO$_2$, which will be negligible in our experiments, due to the low temperature, but may change the ratio between the components in dense clouds.\\

\section*{Acknowledgements}
Authors acknowledge INTA for the formation grant TS 23/17. Additionally, to the Spanish Ministry of Science, Innovation and Universities for the research grant number AYA2014-60585-P and AYA2017-85322-R (AEI/FEDER, UE) and the predoctoral grant FPU17/03172. Finally, to the Ministry of Science and Technology of Taiwan (MOST) for the MOST grants 107-2112-M-008-016-MY3 (Y-JC) and a summer grant.\\




\bibliographystyle{mnras}
\bibliography{bibliography} 








\end{document}